# Enhanced transmission modulation based on dielectric metasurfaces loaded with graphene

Christos Argyropoulos[*]

Department of Electrical and Computer Engineering, University of Nebraska-Lincoln, Lincoln, NE, 68588, USA

*christos.argyropoulos@unl.edu

We present a hybrid graphene/dielectric metasurface design to achieve strong tunable and modulated transmission at near-infrared (near-IR) frequencies. The proposed device is constituted by periodic pairs of asymmetric silicon nanobars placed over a silica substrate. An one-atom-thick graphene sheet is positioned between the nanobars and the substrate. The in-plane electromagnetic fields are highly localized and enhanced with this all-dielectric metasurface due to its zero Ohmic losses at near-IR wavelengths. They strongly interact with graphene and couple to its properties. Sharp Fano transmission is obtained at the resonant frequency of this hybrid all-dielectric metasurface configuration due to the cancelation of the electric and magnetic dipole responses at this frequency point. The properties of the graphene monolayer flake can be adjusted by tuning its Fermi energy or chemical potential, leading to different doping levels and, equivalently, material parameters. As a result, the Q-factor and the Fano resonant transmission spectrum of the proposed hybrid system can be efficiently tuned and controlled due to the strong light–graphene interaction. Higher than 60% modulation in the transmission coefficient is reported at near-IR frequencies. The proposed hybrid graphene/dielectric nanodevice has compact footprint, ultrafast speed, and can be easily integrated to the current CMOS technology. These features would have promising applications to near-IR tunable filters, faster optical interconnects, efficient sensors, switches, and amplitude modulators.

## I. INTRODUCTION

Research on metasurfaces and plasmonic devices is currently shifting towards the incorporation of tunable, nonlinear and reconfigurable materials, which may lead to integrated nanophotonic devices with more pronounced and intriguing functionalities in terms of light manipulation and control [1]. In particular, the efficient concentration of light in subwavelength volumes and the resulting enhanced fields serve as an excellent platform to boost tunability and nonlinear optical processes [2]. Recently, several reconfigurable, tunable and nonlinear plasmonic systems have been presented in the literature. Their operation may be controlled and manipulated with electromagnetic forces [3], phase-change media [4]-[7], liquid crystals [8], electro-optical effects [9]-[10] and optical nonlinearities [11]-[14].

One promising and practical choice among the plethora of recently introduced materials to achieve tunable plasmonic operation is graphene. It is a two-dimensional (2D)—one-atom-thick—conductive material with zero-bandgap and linear dispersion [15]-[16]. It can sustain highly confined surface plasmons in the far- and mid-IR frequencies [17]. Even more interestingly, the properties of graphene can be tuned in the entire IR and optical frequency range by varying its doping level, leading to increased or decreased carrier density concentration inside this ultrathin material. Chemical [16], electrostatic [18]-[19] and, recently, even optical [20] ways to vary the doping level in graphene have been proposed. Tunable surface plasmons propagating on the surface of graphene have already been experimentally demonstrated in the far- and mid-IR frequencies [21]-[28]. Furthermore, the incorporation of graphene in different resonating structures, such as nanoantennas and metamaterial split-ring resonators, has the potential to rapidly increase the interaction between graphene and incident radiation. This may lead to a variety of IR and optical devices with exciting functionalities, such as efficient photodetectors [29]-[30], ultrafast modulators [31]-[34], and tunable absorbers [35]-[37] .

Note that the interband loss mechanism of graphene becomes dominant at near-IR frequencies and surface plasmons cannot be formed and sustained at its surface [17]. However, it is interesting that the properties of graphene can still be electrically controlled [18]-[19] even at these high frequencies, due to the strong variation in its carrier density and, as a result, doping level. Therefore, it is expected that the tunable properties of this 2D material will lead to altered scattering or transmitting responses at near-IR or even optical frequencies when graphene is ingeniously combined with metasurfaces or plasmonic nanostructures. However, graphene can only interact with the tangential (in-plane) electric field components of impinging electromagnetic radiation due to its one-atom thickness.

To this end, metallic nanorod nanoantennas [38] and Fano metasurfaces [39] combined with graphene were employed to experimentally observe electrical tunability of the scattering and reflection response, respectively, at near-IR frequencies. Electrolyte gating made by ionic liquid was used in both cases to control the plasmon or Fano resonance through efficiently tailoring graphene's doping level. However, the fridging in-plane fields in both aforementioned hybrid configurations were relative weak and, as a result, interacted poorly with the graphene region.

This naturally led to relatively weak interaction between graphene and impinging radiation and, as a consequence, poor tunable performance for both hybrid systems. More importantly, metallic structures suffer from inherently high ohmic losses, which further decreased the enhancement of the in-plane electric field components. Therefore, hybrid graphene/nanoantenna and metasurface plasmonic nanostructures have an intrinsically limited tunable operation at the technologically important telecommunication (near-IR) and optical frequency ranges.

On the other hand, subwavelength all-dielectric metasurfaces based on silicon (Si) nanostructures placed over a silica ($SiO_2$) substrate promise zero nonradiative or Ohmic losses at near-IR frequencies and easy integration to the current CMOS technology. The inherent zero-loss properties of these dielectric materials at near-IR can lead to high quality factor (Q-factor) Fano [40] and electromagnetically induced transparency (EIT) [41] resonances. These properties are in stark contrast to the typical plasmonic nanostructures, which are always based on metals and exhibit high Joule losses leading to low Q-factor scattering or transmission responses. Recently, trapped mode magnetic resonances were reported with an asymmetric silicon nanobar configuration to achieve high Q-factor Fano-type transmission and reflection signatures [42]-[43]. In addition, similar dielectric configurations with broken nanobars can also sustain very strong and localized in-plane electric fields in subwavelength regions [44]. It will be interesting to demonstrate a way to obtain reconfigurable and tunable performance with these all-dielectric metasurfaces.

## II. HYBRID GRAPHENE/DIELECTRIC METASURFACE

In this work, we theoretically propose an alternative way to overcome the severe tunability and modulation limitations of hybrid graphene/plasmonic systems. We demonstrate a practical hybrid graphene/dielectric Fano metasurface system. In this configuration, enhanced interaction between graphene and impinging radiation is indeed obtained, thus, allowing for improved sensitivity functionalities compared to previous works mainly based on hybrid plasmonic structures. The proposed device leads to stronger tunability and modulation of the transmission coefficient at the technologically interesting telecommunication (near-IR) wavelengths. The geometry of the proposed hybrid system is based on periodically arranged asymmetric Si nanobars placed over a silica substrate and it is shown in Fig. 1. Silicon and silica are lossless at near-IR frequencies and have constant real refractive indices equal to $n_{Si} = 3.5$ and $n_{SiO_2} = 1.44$, respectively. In the geometry reported in Fig. 1, graphene is embedded between the asymmetric Si nanorods and the silica substrate. It can be epitaxially grown over the silica substrate and the asymmetric nanorods can be placed over the graphene with a regrowth process. However, exactly the same modulation functionalities can be obtained if the graphene is deposited over the dielectric metasurface on top of the silicon nanobars, which can further decrease the experimental complexity of the proposed structure. The dimensions of the silicon nanorods are: $h_1 = 150nm$, $h_2 = 200nm$, $w = 750nm$, $t = 150nm$. They are separated by distance: $d = 250nm$. The unit cell has vertical and horizontal periodicities of: $w_s = h_s = 900nm$. The thickness of the silica substrate is chosen to be: $t_s = 600nm$, but it may be smaller without affecting the performance of the proposed hybrid device.

First, we simulate the all-dielectric metasurface shown in Fig. 1 without the presence of graphene, in order to verify the transmission performance and field distribution of this metasurface. The structure is accurately modelled with full-wave simulations based on the finite integration technique [45] and the transmission coefficient is plotted in Fig. 2(a). An ultrasharp high Q-factor Fano resonance is obtained in the transmission coefficient spectrum due to the trapped magnetic mode at the resonance of this periodic asymmetric silicon nanobar configuration [42]-[43]. The electric field distribution at the resonance transmission dip ($\lambda = 1.59 \mu m$) is also computed by full-wave simulations [45] and reported in Fig. 2(b). Strong antiparallel in-plane electric fields are dominant along the two asymmetric nanobars, which cause cancelation of the electric and magnetic dipole responses at the resonance dip due to destructive interference. This effect reduces radiation losses and, equivalently, coupling to the surrounding free space. It leads to the ultrasharp Fano transmission response. More importantly, the computed in-plane fields are boosted (90 times enhancement compared to incident wave) and confined around and inside the silicon nanobars, as it can be seen in Fig. 2(b). The fields can be further increased if we decrease the periodicity of the unit cell and the neighboring nanorods approach each other [44]. However, the experimental verification of the metasurface will be more challenging due to the strict fabrication tolerances in the manufacturing of silicon nanostructures. The sharp transmission response combined with enlarged fields are ideal conditions to achieve stronger transmission sensitivity when different materials surround this structure or are embedded in this configuration.

## III. GRAPHENE PROPERTIES

In the following, we will present a method to turn the proposed nanodevice to an efficient electrically-controlled transmission modulator. This is possible with the incorporation of one graphene monolayer between the silicon nanobars and the silica substrate, as it is shown in Fig. 1. Graphene's thickness is assumed to be $g = 1 nm$, in agreement to several previous studies of hybrid graphene/plasmonic devices, where full-wave simulations were used to verify the experimental results [38]-[39]. The real and imaginary part of graphene's surface complex permittivity is given, as a function of incident radiation's energy and Fermi level (chemical potential), by the following formulas [38], [46]:

$$\varepsilon_g^{real}(E) = 1 + \frac{e^2}{8\pi E \varepsilon_0 g} \ln \frac{(E + 2|E_F|)^2 + \Gamma^2}{(E - 2|E_F|)^2 + \Gamma^2} - \frac{e^2 |E_F|}{\pi \varepsilon_0 g \left[E^2 + (1/\tau)^2\right]} \qquad (1)$$

$$\varepsilon_g^{imag}(E) = \frac{e^2}{4 E \varepsilon_0 g}\left[1 + \frac{1}{\pi}\left(\tan^{-1}\frac{E - 2|E_F|}{\Gamma} - \tan^{-1}\frac{E + 2|E_F|}{\Gamma}\right)\right] + \frac{e^2 |E_F|}{\pi \tau E \varepsilon_0 g \left[E^2 + (1/\tau)^2\right]} \qquad (2)$$

where $g = 1 nm$ is the thickness of graphene. Parameter $\Gamma$ leads to the interband transition broadening at near-IR frequencies and is taken equal to $\Gamma = 110 meV$, as it has been derived before for the same frequency range by measuring the reflection spectrum of bare graphene flakes [38]. The free carrier scattering rate $1/\tau$ is assumed to be equal to zero because interband transitions dominate at near-IR and optical frequencies over the much weaker intraband transitions [38]. Note that graphene is modeled as anisotropic medium due to its one-atom thickness. The in-plane permittivities are given by Eqs. (1), (2) ($\varepsilon_x = \varepsilon_y = \varepsilon_g$) and the out-of-plane permittivity is equal to free space ($\varepsilon_z = 1$).

We substitute the Planck-Einstein relation $E = hc/\lambda$ ($h$ Planck's constant, $c$ speed of light, $\lambda$ wavelength of radiation) in Eqs. (1) and (2) and plot in Fig. 3 the real [Fig. 3(a)] and imaginary [Fig. 3(b)] parts of graphene's permittivity at near-IR frequencies. The complex permittivity of graphene is plotted in Fig. 3, not only as a function of the impinging radiation's wavelength, but also versus the Fermi energy $E_F$. The Fermi energy $E_F$ is an indicator of the doping level or the increased carrier density inside graphene, and is also called chemical potential. It is usually altered with electrostatic gating and can typically be tuned from -1 eV to 1 eV [47]. Electron doping in graphene is obtained when positive Fermi energy is applied. Otherwise, hole doping is dominant when negative Fermi energy is employed.

Interband transitions prevail at near-IR frequencies when undoped graphene is considered ($E_F = 0 eV$). This can be clearly seen in Fig. 3(b), where the imaginary part of graphene's complex permittivity is particularly high at $E_F = 0 eV$. Hence, graphene acts as a pure resistor in the undoped case. This property leads to the characteristic 2.3% absorption of graphene in optical frequencies, which is exceptionally high considering its ultra-small thickness [48]. However, several interband transition channels are blocked and the absorption of graphene rapidly decreases when the graphene monolayer is doped [by electrons ($E_F > 0 eV$) in our case]. This is clearly captured in Fig. 3(b), where the imaginary part of permittivity drastically reduces for higher doping levels. This exceptional property of graphene combined with the strong coupling to the enhanced in-plane fields due to the all-dielectric metasurface will lead to strong tunability in the near-IR transmission spectrum of the proposed hybrid device.

## IV. STRONG TRANSMISSION MODULATION

The hybrid graphene/dielectric metasurface structure shown in Fig. 1, with dimensions presented before, is also modeled with accurate full-wave simulations [45]. The properties of graphene were mentioned before [Eqs. (1) and (2)] and were plotted in Fig. 3. Normally incident x-polarized (along the Si nanorods) radiation illuminates the proposed hybrid device composed of periodically arranged asymmetric Si nanobars placed over a graphene monolayer and the silica substrate. The transmission coefficient is computed and shown in Fig. 4 for three different doping levels of graphene (Fermi energies: $0eV$, $0.5eV$, $0.75eV$). The pronounced dip in the transmission coefficient Fano resonance, reported before in Fig. 2(a) when graphene was absent, is increased and becomes more flattened in the case of undoped graphene $(E_F = 0eV)$. Furthermore, the sharp resonance is broadened and the Q-factor is rapidly decreased. Note that the in-plane electric fields are strongly coupled to the interband absorption of the undoped graphene monolayer and this leads to the shrunk outline of the Fano resonance. The in-plane fields are enhanced at the resonant wavelength $(\lambda = 1.59 \mu m)$, with a maximum enhancement of 90 compared to the incident radiation, as it can be seen in Fig. 2(b). Similar strong field enhancement has been obtained before in [44]. The enhanced fields are dominant around the silicon nanobars and will couple in the same strong way to graphene even if it is deposited over and not beneath the Si nanorods. This configuration will simplify the potential experimental design of the proposed hybrid device.

To further demonstrate the potential of efficient tunability and modulation in the transmission spectrum of this hybrid graphene/dielectric metasurface system, we weakly vary the Fermi energy or chemical potential of graphene and the results are reported in Fig. 4. Strong amplitude modulation in the Fano transmission is obtained when the Fermi energy is equal to $E_F = 0.5eV$, which becomes even stronger at $E_F = 0.75eV$. This interesting property is combined with an increased Q-factor of the resonant signature because the interband transitions loss mechanism of graphene is rapidly decreased, as the doping level is increased [see Fig. 3(b)]. The robust modulation in transmission of the presented hybrid system is due to the ultra-strong interaction between the impinging radiation and the dominant in-plane properties $(\varepsilon_x = \varepsilon_y = \varepsilon_g)$ of the graphene monolayer. Ultimately, it is based on the altered properties of graphene, which strongly interact with the intense in-plane electric fields of the trapped magnetic mode shown in Fig. 2(b). Note that this effect is not based on the formation of additional surface plasmons propagating on the surface of graphene, as they were used before to create mid- and far- IR tunable devices [22]-[23], [35]-[37]. Surface plasmons along graphene cannot be sustained at near-IR and optical frequencies [17].

The transmission amplitude modulation of the proposed structure can be presented in a more quantitative and clear way by calculating the difference in transmission between heavily doped $(E_F = 0.75eV)$ and undoped $(E_F = 0eV)$ graphene. The transmission difference relationship $\Delta T = T(E_F = 0.75eV) - T(E_F = 0eV)$ is plotted in Fig. 5 as a function of the impinging's radiation wavelength. Interestingly, the transmission difference $\Delta T$ can reach values higher than 60% at the Fano resonance transmission dip $(\lambda = 1.59\mu m)$. High transmission modulation is obtained at this particular wavelength. Note that moderate modulation is also obtained within a narrow wavelength range around $\lambda = 1.59\mu m$ [Fig. 5]. These results can lead to the design of an efficient and tunable amplitude modulator in the near-IR frequency range and can have interesting biosensing and telecommunication applications.

Several approaches have been proposed to increase the carrier concentration or doping level in graphene nanostructures. In the proposed hybrid configuration, one potential approach will be to use transparent electrodes placed between either the silica substrate or the silicon nanorods and the graphene. These electrodes can be used to electrostatically gate graphene and increase the Fermi energy in the current configuration. The Fermi energy is given by the formula $E_F = \hbar v_F \sqrt{\pi C V_g / e}$ [15], where $\hbar$ is the Planck's constant, $v_F = 10^6 m/s$ the Fermi velocity, $V_g$ the applied gate voltage between the electrodes and $e$ the electron's charge. The electrostatic gate capacitance $C$ depends on the distance and material between the electrodes. If the thickness of the silica substrate is reduced to $t_s = 100nm$, the capacitance $C$ will increase but the Fano transmission will be unaffected (not shown here). This resonant response is generated by the silicon metasurface and it is independent of the substrate's thickness. In this scenario, the transparent electrodes can be placed between the silica substrate and the graphene, which will lead to an approximate gate voltage of $V_g \simeq 4.5V$ in order to obtain Fermi energy equal to $E_F = 0.75eV$. In an alternative design, we can place the transparent electrodes between the silicon nanorods and the graphene. In this case, even lower gate voltage $V_g \simeq 2.3V$ will be needed to obtain $E_F = 0.75eV$ because of the higher static permittivity of silicon compared to silica. Hence, realistic values of gating voltage are required to create an efficient electrostatic transmission modulator with the proposed hybrid nanostructure.

Another promising approach to change the Fermi energy is the electrochemical gating based on ionic liquid electrodes [49], which was used before to gate other hybrid graphene/plasmonic structures [38]-[39]. In addition, one more interesting alternative approach is the recently introduced photo-induced doping in graphene [20]. It will provide exceptional high doping levels, in particular for the currently proposed design, due to the enhanced in-plane fields along the surface of graphene. We expect that the current theoretical study will stimulate increased interest towards the experimental verification of the proposed hybrid transmission modulator. Moreover, the dimensions of the dielectric metasurface can be scaled down to achieve Fano transmission resonances at visible wavelengths. However, silicon losses will be larger at visible frequencies due to its band-gap and the ultrasharp Fano resonance will be broader compared to the currently presented results. Finally, it is worth mentioning that another very interesting potential application of the proposed hybrid system is to enhance several optical nonlinear processes [12]. It is very encouraging that graphene has been recently found to possess extremely high nonlinear coefficients [50]-[51]. This is another exciting functionality of the proposed device that is envisioned to lead to several new nanophotonic applications.

## V. CONCLUSIONS

To conclude, we have presented an ultracompact hybrid graphene/dielectric metasurface device, which can achieve unprecedented transmission amplitude modulation in the technologically interesting telecom (near-IR) wavelength range. The complex surface permittivity of graphene was demonstrated as a function of Fermi energy and wavelength of operation. Graphene was placed over silica and its properties were strongly coupled to the in-plane fields of the proposed all-dielectric silicon metasurface design. This metasurface does not suffer from Ohmic losses and is advantageous compared to usual plasmonic metasurfaces. The resulted ultrasharp Fano resonance in the proposed configuration is based on the trapped magnetic mode and was found to exhibit remarkably strong coupling with graphene. It can be switched on or off depending on the properties of graphene. This is in plain contrast to relative weak coupling reported in previous works related to hybrid near-IR graphene/plasmonic systems [38]-[39]. In addition, the doping level of graphene and the required gating voltage had practical and relative low values, i.e. low energy consumption is needed to achieve the reported tunable performance. The operational principle was based on tuning the complex surface permittivity of graphene by applying external voltage or, equivalently, increasing doping level, which led to efficient amplitude modulation in the transmitted radiation. Interestingly, different all-dielectric structures, such as dielectric waveguides [52]-[53], photonic crystals [54] or microring resonators [55], coupled to graphene have been recently proposed to induce strong transmission or reflection modulation.

The currently proposed device provides an ideal and unique platform to tame, modulate and control the amplitude of transmission coefficient with all-dielectric metasurfaces. It has a particular small device footprint, ultrafast response and low power consumption. For example, 50 by 50 unit cells of the proposed metasurface are enough to obtain an efficient transmission modulator operating over an ultra-compact area $A = 45 \times 45 \mu m^2$ with an estimated capacitance $C = A\varepsilon/t_s \simeq 0.7 pF$, where $\varepsilon = 4$ is the static permittivity of silica and the gate separation is assumed $t_s = 100 nm$. Ultimately, the speed of the proposed device is limited by the surface resistance of graphene $R_s \simeq 20 \Omega$ [55] and it is approximate equal to $1/2\pi R_s C \simeq 11 GHz$. In addition, the sharp characteristics of the presented Fano resonance has the advantage of enabling large switching transmission contrast and may even further accelerate the modulation speed due to the unique high Q-factor response. Several future integrated nanophotonic components and optoelectronic systems are envisioned based on the proposed hybrid nanostructure, such as ultrafast photodetectors with enhanced responsivity, new biosensors and efficient electro-optical modulators. Similar all-dielectric metasurfaces may also be combined with other 2D materials (MoS$_2$, hBN) or Van de Waals heterostructures to achieve even more functionalities.

**ACKNOWLEDGMENTS**

This work was partially supported by the Office of Research and Economic Development at University of Nebraska-Lincoln.

**Figures**

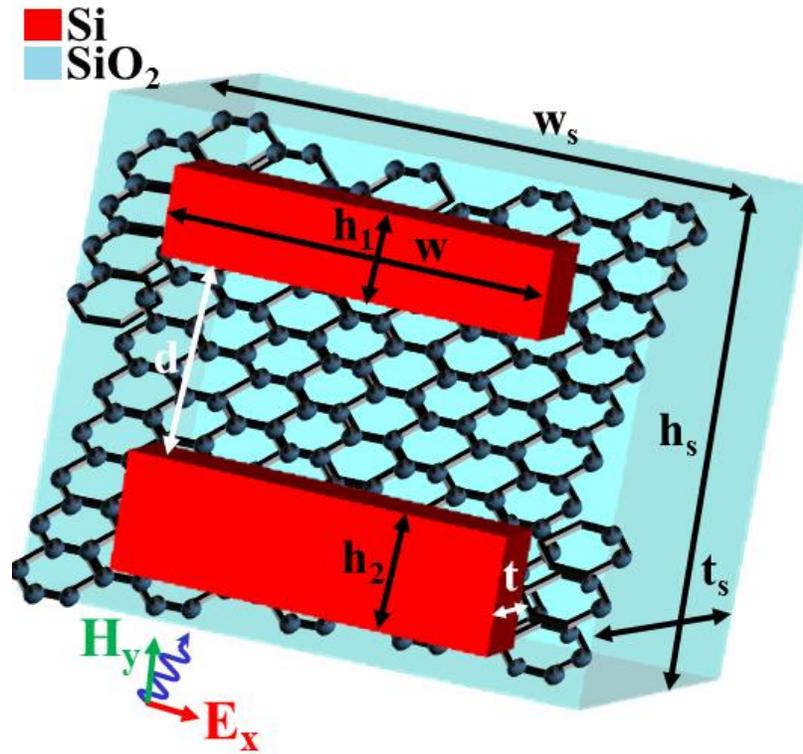

**Figure 1** Schematic of the hybrid graphene/dielectric metasurface. The proposed structure is composed of a periodically arranged pair of asymmetric silicon nanobars placed over a silica substrate. Graphene is placed between the nanobars and the substrate. The dimensions of the hybrid system are: $w = 750nm$, $t = 150nm$, $h_1 = 150nm$, $h_2 = 200nm$, $w_s = h_s = 900nm$, $t_s = 600nm$, and $d = 250nm$.

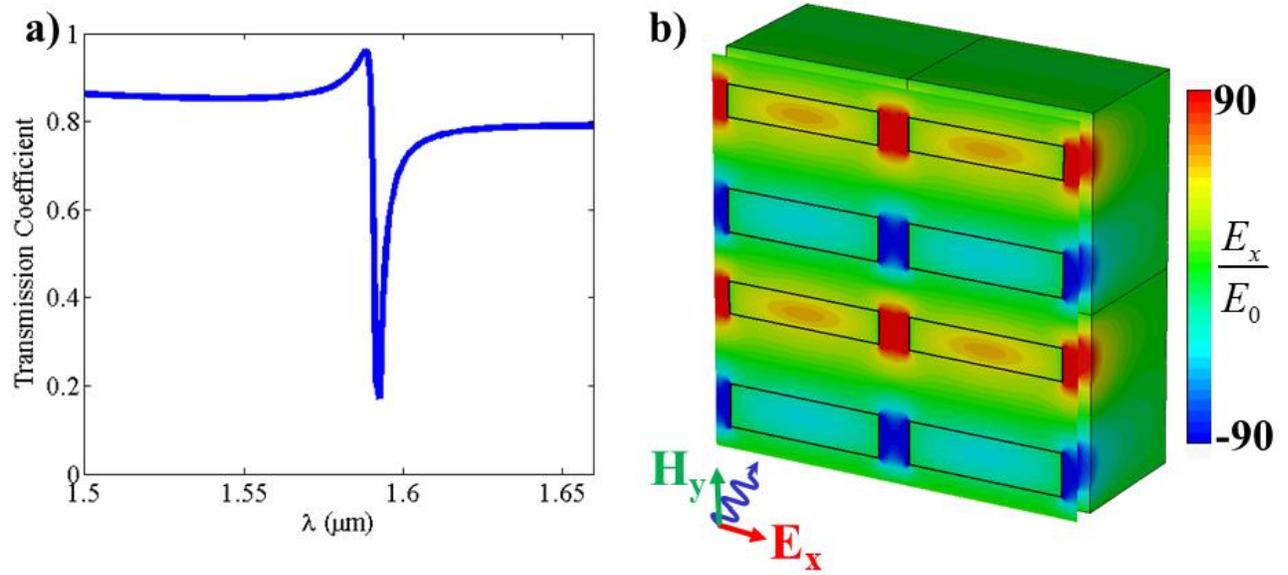

**Figure 2** (a) Fano resonance obtained in the transmission coefficient of the proposed all-dielectric metasurface shown in Fig. 1. (b) Simulated in-plane electric field ($E_x$) distribution monitored at the dip of transmission coefficient ($\lambda = 1.59\,\mu m$). Enhanced in-plane fields are obtained and their antiparallel formation along the two asymmetric nanobars leads to the magnetic Fano resonance shown in caption (a). Graphene is removed from the calculations presented in this figure.

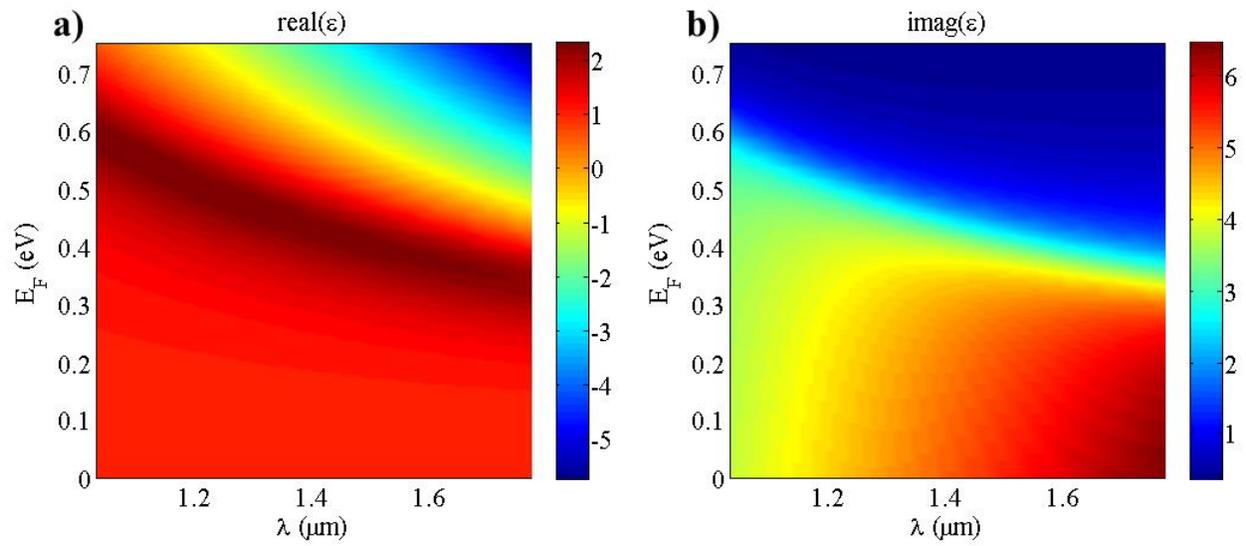

**Figure 3** (a) Real and (b) imaginary part of the complex permittivity of graphene as a function of the impinging radiation's wavelength and the Fermi energy. As the Fermi energy or doping level increases, the loss in graphene [imaginary part of permittivity/caption (b)] rapidly decreases.

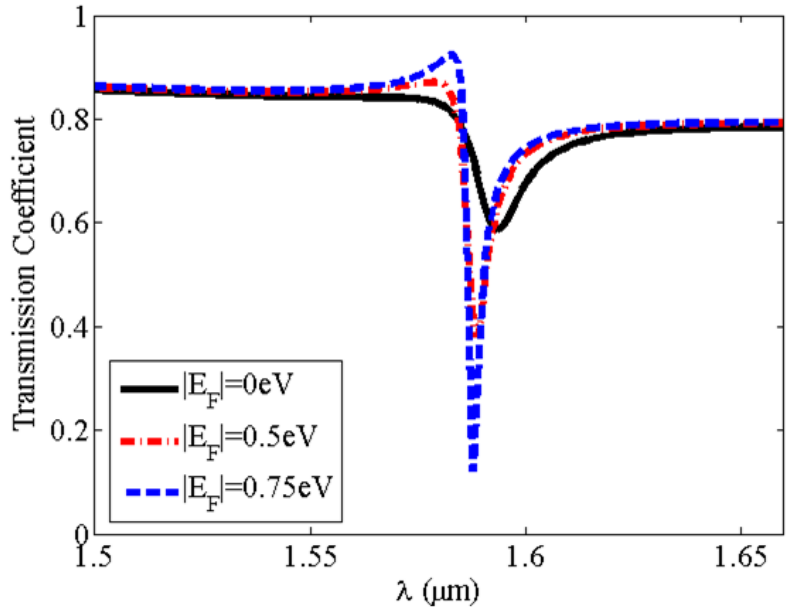

**Figure 4** Transmission coefficient versus incident radiation's wavelength of the proposed hybrid graphene/dielectric metasurface shown in Fig. 1. The transmission excursion and Q-factor of the Fano resonance is increased and strongly modified for higher doping levels of graphene

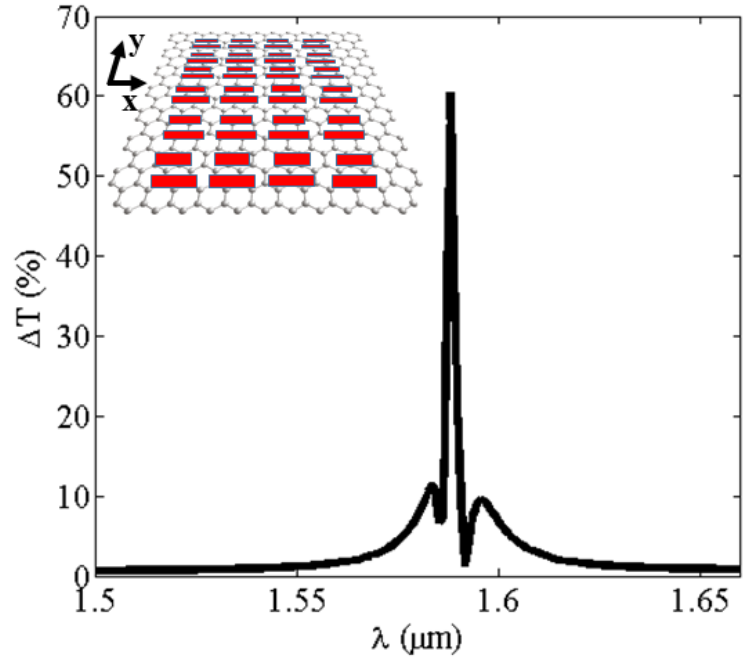

**Figure 5** Computed percentage of transmission coefficient difference versus the incident radiation's wavelength. The properties of the proposed hybrid system change from heavily doped ($E_F = 0.75 eV$) to undoped ($E_F = 0 eV$) graphene. The amplitude of the transmission coefficient can be modulated by more than 60% at $\lambda = 1.59 \mu m$.